# SOLID PROPELLANTS

B. P. MASON,[1,]* C. M. ROLAND[2,]*

[1]DEPARTMENT OF PHYSICS, NAVAL POSTGRADUATE SCHOOL, MONTEREY, CA 93943-5216
[2]CHEMISTRY DIVISION, CODE 6105, NAVAL RESEARCH LABORATORY, WASHINGTON, DC 20375-5342

## ABSTRACT

Solid propellants are energetic materials used to launch and propel rockets and missiles. Although their history dates to the use of black powder more than two millennia ago, greater performance demands and the need for "insensitive munitions" that are resistant to accidental ignition have driven much research and development over the past half-century. The focus of this review is the material aspects of propellants, rather than their performance, with an emphasis on the polymers that serve as binders for oxidizer particles and as fuel for composite propellants. The prevalent modern binders are discussed along with a discussion of the limitations of state-of-the-art modeling of composite motors. [doi:10.5254/rct.19.80456]

## CONTENTS

## I. INTRODUCTION

Solid composite propellants are highly-filled elastomers used prominently as energetic materials for military ordnance and rockets, with commercial applications such as gas generation. All tactical (i.e., battlefield) missiles use solid propellants in some form. Composite propellants are distinguished from double-base propellants,[1] typically composed of nitrocellulose (NC) and nitroglycerin, used in smaller charges for firearms and mortars. The advantages of solid rocket propellants include: (i) simplicity, which is important for maintenance costs and savings in high production rate systems; (ii) storage stability, with service lifetimes that can be as long as 30 years; (iii) resistance to unintended detonation; (iv) reliability, related to their simplicity and chemical stability; and (v) high mass flow rates during launch, and consequently high thrust (propulsion force), a requirement for the initial phase of missiles, all of which use solid propellant boosters. Two disadvantages of solid propellants are the difficulty in varying thrust on demand (i.e., solid fuel rockets generally cannot be throttled or operated in start-stop mode) and relatively low specific impulse (time integral of the thrust per unit weight of propellant), $I_{sp}$, in comparison with liquid fuel motors. These preclude their use as the main propulsion method for commercial satellites and space probes, although solid rocket motors (SRM) have a long history as boosters. They also find application in aircraft using jet-assisted takeoff (JATO) to launch quickly or when overloaded. Initial experiments with gliders date to the 1920s, and SRM were developed for JATOs in World War II for aircraft using short runways and for rudimentary barrage rockets.

*Corresponding authors. Email: brian.mason@nps.edu; roland@nrl.navy.mil

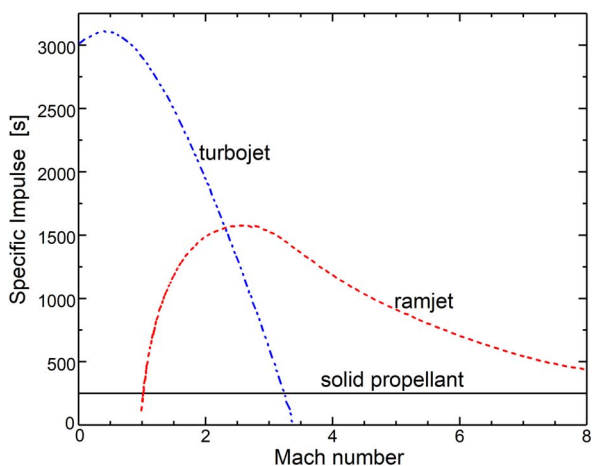

FIG. 1. — Representative values of impulse normalized by fuel weight as a function of speed for three common propulsion systems. Mach number is ratio of speed of missile to that of sound (∼340 m/s, depending on temperature and altitude).[6]

Rocket propellants differ from other fuels by not requiring an external source of oxygen; the oxidizer is a component of the propellant. SRM are by far the most widely used engines for missile propulsion. A related motor type is the ramjet, which typically uses liquid fuel, although solid-fuel ramjets are in limited use. In a ramjet, oxygen is brought in from the surrounding atmosphere.[2] A variation on ramjets is the turbojet motor, which uses a compressor to enhance air speed, allowing subsonic operation. Hybrid rocket motors combine solid and liquid propellants, affording throttling and start-stop capabilities lacking in conventional SRM.[3] Other systems include liquid propellants, used primarily for space applications, and nuclear motors, which offer enormous energy densities and the largest specific impulse of any engine. Nuclear systems have been limited to surface ships and submarines, although very recently there have been unverified claims of nuclear-propelled missiles and underwater drones.[4,5]

The performances of the propulsion systems employing inert binders are compared in Figure 1 (adapted from ref 6), which shows representative specific impulse values versus missile speed. Specific impulse is a measure of motor efficiency, and for solid propellants, this depends on operating conditions, such as the combustion chamber pressure, nozzle expansion rate, and the ambient pressure. The $I_{sp}$ of the solid propellant becomes constant within a few milliseconds of ignition.

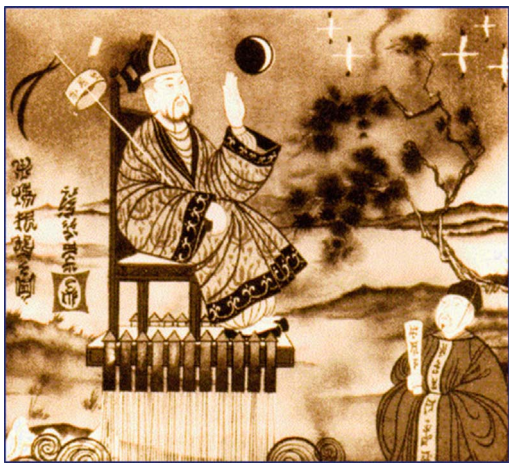

FIG. 2. — Wan Hu was the first recorded astronaut (*ca*. 1500 AD). Forty-seven solid rocket motors based on potassium nitrate, sulfur, and charcoal contained in bamboo tubes were simultaneously lit, resulting in a successful launch or a catastrophic explosion, depending on the account. In neither version was Hu ever seen again. (Image courtesy of L. T. DeLuca, Politecnico di Milano.)

## II. BACKGROUND

The earliest chemical explosives, dating to the first millennium AD, were based on gunpowder ("black powder"), a mixture of potassium nitrate (saltpeter), sulfur, and charcoal (or less often coal). Used for fireworks (Chinese "fire drug"), ordnance-employed gunpowder was limited to moderate propellant grain sizes, because of the difficulty in compressing large quantities without introducing cracks or holes that caused erratic combustion. Although the story is perhaps apocryphal, the first astronaut was reputed to have used solid propellants (Figure 2). (The launch vehicles for the first Soviet and American astronauts employed liquid fuel engines.) A major advance in the late 19th century was development of "smokeless powder," based on NC, nitroglycerin, and/or nitroguanidine. The term derives from the low concentration of particulates, and thus smoke, among the combustion products. Many developments ensued, mainly directed to improving the stability and reliability of smokeless powder. Note that outside of the United States, smokeless powder is referred to simply as *propellant*, a generic term for proprietary versions such as Ballistite and Cordite.

Presently, solid propellants are used for the launch systems of many civilian and military rockets,[7] mainly because of their greater safety and reliability in comparison with liquid fuel. Early booster charges were relatively small (<30 kg); in comparison, each booster on the Space Shuttle had 500 000 kg of solid propellant. The thrust-to-weight ratio is a dimensionless quantity that indicates a vehicle's acceleration capability. For a solid propellant booster rocket, it can exceed a ratio of 100, which is two orders of magnitude greater than for supersonic fighter aircraft. The largest SRM were the two booster rockets on NASA's Space Launch System (SLS). Each booster burned six tons of poly(butadiene-acrylonitrile)/ammonium perchlorate (AP) propellant per second, achieving a combined maximum thrust of almost 40 MN. The design of the SLS included three additional SRM systems: a jettison motor, an abort system, and an attitude control motor. Other systems using solid propellant motors as boosters and retrorockets (used for deceleration and turning) include the Atlas V and Delta IV medium+ rockets. Solid propellants also serve as the primary thrust system in missile defense systems such as the Aegis and Patriot missiles and the Minuteman III ICBM. The European commercial launch vehicles Ariane and Vega employ multiple solid rocket boosters.

TABLE I
OXIDIZER PROPERTIES[13,a]

| | AP | AN | RDX | HMX | CL-20 | AND | HNF |
|---|---|---|---|---|---|---|---|
| Molecular weight, Da | 118 | 80 | 222 | 296 | 438 | 124 | 183 |
| Density, g/mL | 1.95 | 1.72 | 1.81 | 1.91 | 2.04 | 1.81 | 1.86 |
| Heat of formation,[b] kJ/g | −2.51 | −4.95 | 0.325 | 0.28 | 0.85 | −1.21 | −0.39 |
| Oxygen balance,[c] % | 34 | 20 | −22 | −22 | −11 | 26 | 13 |
| Impact sensitivity,[d] cm | 15 | >49 | 7.5 | 7.4 | 2.5 | 3.7 | 3 |
| Friction sensitivity, N | >100 | 350 | 120 | 120 | 124 | >350 | 20 |

[a] AP, ammonium perchlorate; AN, ammonium nitrate; RDX, Royal Demolition Explosive: cyclotrimethylenetrinitramine; HMX, Her Majesty's Explosive: cyclotetramethylenetetranitramine; CL-20, China Lake 20: 2,4,6,8,10,12-hexanitro-2,4,6,8,10,12-hexaazaisowurtzitane; HNF, hydrazinium nitroformate.
[b] Negative value indicates exotherm.
[c] Ratio of oxygen in a material to amount required for its complete oxidation.
[d] Drop height of arbitrary weight for which explosion induced.

The most common application of solid propellant is automobile airbags, although the propellant is neither a composite nor a polymer. A variety of compounds, including sodium azide mixtures, nitroguanidine, and tetra- and triazoles, are used. In response to various sensors monitoring acceleration, impact, wheel speed, and so forth, electrical ignition of the propellant causes gas production that inflates the airbag. The time from vehicle impact to full deployment is less than 0.1 s. The largest recall in automotive industry history occurred when ammonium nitrate (AN) was used as the propellant. Exposure to moisture or heat destabilizes AN, with its subsequent reaction transpiring too fast. Rather than inflating the airbag, the rapid gas evolution shatters the steel case containing the airbag. About 70 million vehicles in the United States were affected by the recall, with 20 deaths worldwide attributed to the faulty airbags.

## III. INTERNAL AERODYNAMICS OF SRMs

Considerable effort is devoted to developing accurate descriptions of the internal aerodynamics of SRM, with the objective of predicting operation during ignition, steady-state operation, and termination upon exhaustion of the fuel. Particularly in the unsteady regimes, during which the pressure changes strongly with time, a detailed analysis of flow inside the combustion chamber is required. Even during steady-state burning, which occupies most of the operating time, the propellant burn rate can be affected by pressure, temperature, and gas flow rate. Important issues are motor stability, the relationship of pressure to thrust, and the velocity of the combustion products. The chemistry is complex,[8] involving flowing reactants that reach temperatures that can exceed 3000 K, at pressures as high as 5 MPa (for large booster rockets). Because the combustion area varies during flight, thermodynamic conditions are constantly changing. The severity of these conditions precludes detailed measurements inside the combustion chamber; usually only exit pressures are available. Modeling involves substantial computation times and is hindered by the accuracy of input data and uncertainties regarding the effects of flow turbulence, interactions between metal droplets, unsteady propellant combustion, vibration of motor components, and so

SOLID PROPELLANTS 5

TABLE II
REPRESENTATIVE HTPB-BASED SOLID PROPELLANT FORMULATION

| Component | Amount, % |
|---|---|
| Polymer | 8–10 |
| Isocyanate | 1–2 |
| Plasticizer | 2–10 |
| Oxidizer | 60–85 |
| Metal | 0–20 |
| Bonding agent | 1 |
| Burn rate modifier | 1 |
| Antioxidant | 0.1 |
| Catalyst | 0.1 |

forth. For these reasons, calculations intended to optimize motor design are semiempirical and rarely can be quantitatively validated.

## IV. PERFORMANCE OF SOLID ROCKET PROPELLANTS

A rocket is defined as a device or vehicle for which the accelerating force (thrust) is achieved by expelling mass. Strictly speaking, a rocket is a means of propulsion, distinguished from a missile, which is something that is propelled (for example, by a rocket). The essential components of SRM are the combustion chamber and its exit port (the nozzle), the igniter, and the propellant. Typically, the propellant consists of oxidizing particles (e.g., AP or AN) embedded in a fuel, which includes both a polymeric binder and metal powder. Light metals such as aluminum are used; their function is to enhance the degree of combustion of the binder.[9,10] The particles comprising the propellant are referred to as the grain. In practice, most solid SRM use nonenergetic binders, which are polymers that require an oxidizer to burn.

By far, the most popular inorganic oxidizer is AP. It produces completely gaseous products, an advantage over potassium nitrate and potassium perchlorate. AN is an alternative to AP but is less energetic and has a slower burn rate.[11] Dual-oxidizer systems, such as AP or AN with ammonium dinitramide (ADN), can offer advantages, such as enhanced specific impulse and lower HCl production.[12] The properties of various oxidizers relevant to propellants are compared in Table I.[13]

The use of oxidizers distinguishes composite propellants from inherently energetic materials such as NC.[14] When conventional energetic binders cannot provide the required structural integrity for a particular motor application, inert polymers, such as polyurethanes and polyethers, can be made energetic by introduction of reactive groups.[15]

Binders based on polymer networks have superior mechanical properties, with the crosslinking providing shape stability and resistance to cracking and void formation, while also serving as fuel. The amount of binder is determined by the size of the combustion chamber. Upon propellant ignition, combustion products form that are emitted through the nozzle. The thrust for a given motor can be varied by changing the composition of the energetic materials or their burn rates, the latter by changing, for example, the nozzle geometry. Early inert binders were blends of a high concentration of potassium perchlorate with asphalt (see discussion below). The mechanical properties were poor, and these have been replaced entirely with formulations based on polymer networks, most commonly polybutadiene with functional end-groups. An example is hydroxyl-terminated polybutadiene (HTPB), which is reacted with isocyanates or epoxies to form networks. By design, the crosslinking agent is usually below stoichiometric levels to ensure a significant soluble fraction of polymer remains after network formation.

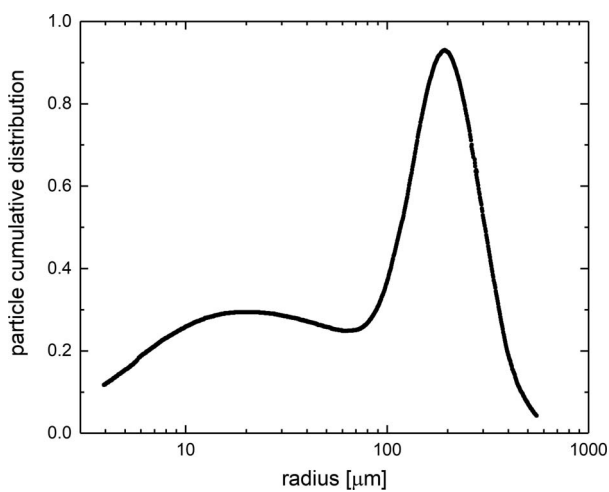

FIG. 3. — Cumulative distribution of particles in a representative grain.[27]

The main requirement of the fuel is that its oxidation is strongly exothermic and accompanied by the production of gaseous products, with the heat and kinetic energy of the effluents serving as the propulsion mechanism. The relation describing this is the classic rocket equation obtained from conservation of momentum and Newton's second law:[6]

$$v_f - v_i = v_e \ln \frac{m_i}{m_f} \qquad (1)$$

where $v$ is the rocket velocity and $m$ its mass, with subscripts $i, f,$ and $e$ denoting initial conditions, post burnout, and the exhaust velocity, respectively.

The principal parameters in the engineering of a solid propellant are the burn rate and grain design (e.g., particle size and shape).[7] Except during ignition or when instabilities arise, the burn rate of conventional SRM is essentially constant (Figure 1) and depends only on the initial temperature and pressure. In pulsed rocket motors, the solid propellant is partitioned into multiple sections.[16] Propellant in the forward segment burns quickly, providing rapid acceleration, with the remaining segments ignited on command. The slower rates of acceleration reduce stresses on the rocket, and the partitioning provides control, including the ability to stop and subsequently reignite the motor. Another method to control burn rate[17] is to incorporate a surfactant into the grain that causes the combustion of the propellant to depend on pressure. At a sufficiently high pressure, the combustion of the propellant is extinguished. Pressure affects combustion rates because of its effect on the gas phase above the combustion surface.[18,19] Ideally, only a thin layer of propellant at the surface combusts, so that the grain temperature prior to combustion remains close to the external temperature.

A typical solid propellant formulation is shown in Table II. Filler volume content is typically in the range of $\phi = 70$ to 80%, although concentrations as high as 90% have been used. Higher levels of solids loading obviously makes processing and casting of the propellant mixture more difficult. For the very high particle concentrations relevant to solid propellants, the viscosity, $\eta$, of the precured binder is critical. It must be low enough to allow processing but sufficiently high to facilitate dispersion of the particles.[20] The processing problem can be overcome to some extent by using a blend of small and large particles, with small particles occupying the interstitial regions around larger particles. Thus, multimodal size distributions are used in solid propellants to increase the packing fraction, otherwise limited to ∼64% by volume for a uniform particle distribution.[21] Having large particles 10-fold greater in size than the smaller particles enables the mixture to be treated as a suspension of the former in fluid containing the latter. The details of the particle distribution are important, potentially affecting binder adhesion, "hot-spot" formation during combustion, and agglomerate uptake by the gas flow.[22] Both modeling and experiments have been brought to bear to characterize the particular structure in solid propellants.[23–26] Figure 3 shows a representative size distribution of the HTPB solid propellant with $\phi = 0.76$ of AP and aluminum flake.[27]

The maximum amount of propellant yields the best motor performance; however, grain design reflects a compromise between the maximum packing, $\phi_m$, and the desired thrust. A greater propellant content increases the bore diameter and ignition area at the nozzle entrance, critical factors in achieving the desired thrust.

## V. POLYMERIC BINDERS

The polymer in a solid propellant functions as both the binder for the ingredients of the grain and as a fuel. The first requires a polymer of sufficient strength, but it should have low viscosity prior to curing (crosslinking or chain extension) to facilitate mixing and casting. An obvious requirement of any fuel is a high specific energy yield, and for propellants, the combustion products must be gaseous. Particularly for tactical rockets, in which the propellant is exposed to varying ambient temperatures, an insensitivity of burn rate to temperature and pressure is advantageous. Other requirements include storage stability and good adhesion to the oxidizer and other filler particles and (for case-bonded propellants) to the liner. Note that the compositions of SRM binders are similar to those used for plastic-bonded explosive warheads.[28] The most historically important polymers for modern propellant applications are discussed below.

### A. NITROCELLULOSE

SRM using NC occupy an important place in the history of propellants. NC is technically the first modern SRM polymer binder, used initially in nearly pure form (single base) or plasticized with nitroglycerin (double base) for small arms and larger bore weapons in the late 19th century.[29] Early NC-based barrage rockets of the Second World War used essentially the same extruded powders of their gun counterparts. Although accuracy was poor, NC-based rockets were well-regarded when used for concentrated artillery bombardment because of their psychological effect on the enemy.[30]

It is worth noting that while the elastomeric polymer binders discussed below were in development, parallel efforts using NC-based composite rocket motors reached a high level of maturity. These motors used NC "casting powders," a simple mixture of NC and various solid ingredients, extruded and cut to form right circular cylinders of roughly 1 mm. The powder was then loaded into a mold (often the bond-lined rocket case itself, or a free-standing grain subsequently cartridge loaded), with interstitial space filled with nitroglycerin and other plasticizers. Done competently, interdiffusion of the solid and liquid ingredients created a single monolithic rocket motor grain. Such motors were used for the Scout family of rockets and the Naval Research Laboratory's Project Vanguard.[31]

Like their elastomer-based counterparts, double-base rocket motor technology culminated in composite-modified double-base (CMDB) propellants, which contained aluminum, AP, and on occasion high explosives like Her Majesty's Explosive (HMX). CMDB is typically used in ballistic missile motors, such as the third stage of the Minuteman I missile and the second stage of the Polaris A2/A3 missiles.[32]

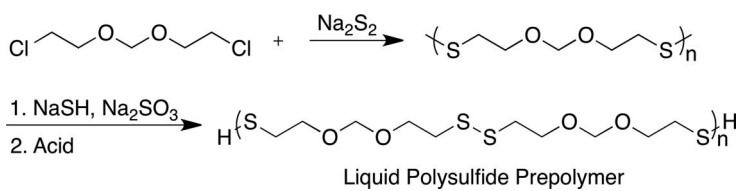

SCHEME 1. — Synthesis of polysulfide prepolymer.

Double-base formulations incorporating NC and NG are not obsolete; they are low cost, reliable, easy to ignite, and composed of readily available ingredients. Double-base motors are still in production and used in current U.S.-made tactical rocket motors, often in the form of a "carpet roll," that is, a sheet of double base that has been rolled around a removable mandrel to form a cylindrical motor having a center perforation.

### B. ASPHALT

Modern elastomer-based propellants are typically traced to the "first" SRM propellant, a mixture of asphalt and potassium perchlorate developed at the California Institute of Technology Gugenheim Aeronautical Laboratory (GALCIT, which subsequently became the Jet Propulsion Laboratory), led by Frank Malina, a graduate student working under the direction of GALCIT director Theodore von Kármán. The goal was to build a rocket that could attain an altitude of 100 000 feet.[33] Discouraged by repeated experimental failures, von Kármán and Malina showed mathematically that a constant ratio between the burning surface area of the solid motor and the nozzle area would result in stable pressure in the rocket body.[34] Using this idea, their initial black powder motors were redesigned, with an unprecedented 12 s of burning with 125 pounds of thrust achieved.[33] However, these motors were overly sensitive to changes in ambient temperature, failing catastrophically in hot weather.

The solution was a new formulation developed by GALCIT member Jack Parsons, consisting of 76% potassium perchlorate oxidizer in a slurry of 7 parts road asphalt binder and 3 parts plasticizing oil.[35] This mixture was heated and cast directly into the motor case without a liner.[33] Although the asphalt–perchlorate formulation was used by the U.S. Navy for JATO rockets, the motors had a number of drawbacks, in particular a limited operational temperature range (ambient ± 30 °C) and low solids loading that resulted in poor mechanical properties.[36] However, this type of motor is usually deemed the first viable SRM, and Parsons has been called the "Father of Rocket Science,"[37] an honorific shared with Werner von Braun, Robert Goddard, and Konstantin Tsiolkovsky.

### C. POLYSULFIDES

In the latter stages of World War II, SRM binder work for JATOs led to the use of a styrene–butadiene rubber, Buna-S, developed by IG Farben in Germany before the war.[38] Although Buna-S performed well over a broad temperature range, dewetting of the motor grain from the case proved an insurmountable problem. Adhesion at the motor's bond line is critical because the debonded surface of the grain will burn prematurely, causing motor failure. By 1945, polychloroprene (CR) binder formulations with potassium perchlorate were developed at the Jet Propulsion Laboratory. These operated over an acceptable temperature range and adhered well to the liner.[34] However, the CR was high molecular weight, and thus processing the binders required large roll mills. The problem was exacerbated by the high loading of the perchlorate solid oxidizer. Various plasticizers were tried, but none adequately improved the processing.

The next step was to replace the CR with a curable liquid prepolymer made by Thiokol Chemical Corporation. Such telechelic prepolymers had been discovered in 1920[39,40] and manufactured by Thiokol since the 1930s. In fact, these liquid rubbers were already in use by the American military for sealing fuel tanks. The first domestic synthetic rubber was Thiokol polysulfide, a widely used sealant, for which the inventor, J. C. Patrick, received the Goodyear Medal in 1958. (Ironically, Thiokol gained notoriety for its role in the Space Shuttle Challenger disaster, in which an SRM booster failed at liftoff due to leakage of a fluoroelastomer O-ring seal.[41]) The initial formulations were based on polythiol prepolymers made by reacting dichloroether with sodium polysulfide (Scheme 1).[42] This rubber could then be treated with sodium hydrosulfide and sodium sulfite to cleave it in a somewhat controlled fashion into liquid prepolymers featuring thiol end groups ($M_W$ = 1000–4000 Da). These oligomers could then be formulated with oxidizers and other ingredients and cured to oxidize the thiols back into disulfides, typically with *p*-quinonedioxime.

Polysulfide-based propellant formulations became the standard for U.S.-made SRM. The technology reached maturity as the fuel for the MGM-29 Sergeant surface-to-surface missile fielded by the U.S. Army in 1961, as well as the second and third stages of the Jupiter-C sounding rocket. The end of polysulfide-based rocket motors came with the discovery in 1955 that aluminum powder added to the standard perchlorate binder significantly increased specific impulse. This discovery could not be applied to polysulfide systems because reactions between the aluminum and binder caused "storage instabilities" (i.e., they exploded).[43]

### D. PLASTISOLS

In 1950, a group at Atlantic Research Corporation (ARC) focused their SRM binder efforts on plastisols of NC and poly(vinyl chloride) (PVC).[44] Similar in concept to NC casting powders, plastisol particulates were much smaller, on the order of 50 microns or less,[45] which allowed them to be suspended in an equal amount of plasticizer and mixed with the oxidizers and metal fuels. One notable quality of PVC plastisol was its long pot life; PVC in a plasticizer such as dibutyl sebacate was stable for years without dissolution.[46] There would be no appreciable hardening of the slurry until it reached 104 °C, and at 150 °C, the mixture would "cure" (that is, the PVC would dissolve in the plasticizer, solidifying the binder in about 5 min). The obvious drawback of such a material was the absence of chemical crosslinks, which limited the working temperature range and prevented bonding to a rocket motor case. Despite these limitations, PVC plastisols found a number of applications in gas generators (e.g., the Polaris A-3 thrust vector control) and in rocket sustainer motors, such as the Mark 30 in ARC's Standard Missile, as well as in the Stinger Launch and the fairing motor for the Trident I C-4.[46] These motors typically showed good aging characteristics and could remain in service for several decades.

Plastisol nitrocellulose (PNC), sometimes called "spheroidal" or "pelletized" nitrocellulose, was developed at ARC and also featured a long pot life, although not as long as PVC's. PNC used NC nitrated at 12.6%, dissolved in nitromethane and ethyl centralite, and emulsified in water. The nitromethane was then leached out and 5–50 micron spheres collected by centrifugation.[47] While used sparingly in SRM, PNC was employed in warheads and gun propellants. The U.S. Navy deemed PNC important enough that it manufactured the material itself at the Naval Propellant Plant in Indian Head, Maryland, starting in 1958. Production ended when accidents forced closure of the facility in the 1990s.[48]

### E. POLY(BUTADIENE-ACRYLIC ACID)

In the mid-1950s, butadiene-containing copolymers reappeared with poly(butadiene-acrylic acid; PBAA), a random copolymer developed as a binder by Thiokol at Redstone Arsenal.[43] PBAA had a molecular weight of about 3000 Da, but being made by free radical emulsion polymerization, it was a mixture of various polyfunctional chains, including some nonfunctional, resulting in poor reproducibility. Crosslinking was carried out by ring opening of difunctional epoxides and aziridines by the acrylic acid groups.[49] Although the prepolymer had a sufficiently low viscosity to allow high solids loading, the resulting crosslinked polymer had poor mechanical properties because of the irregular network structure and for this reason was abandoned. However, PBAA was used by Thiokol as the binder for the first stage of the Minuteman I ICBM.[32]

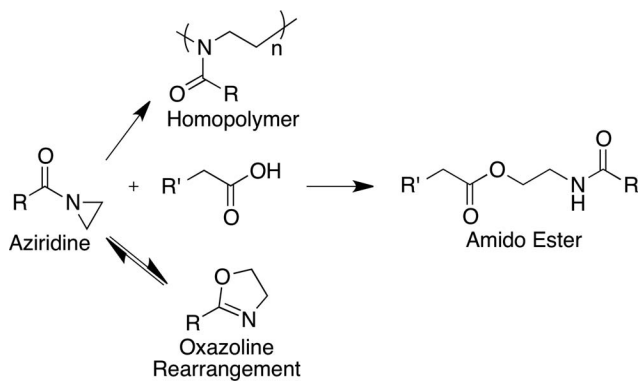

SCHEME 2. — Side reactions in aziridine-based curing.

### F. POLY(BUTADIENE-ACRYLONITRILE-ACRYLIC ACID)

Because of the difficulties in reproducibly crosslinking PBAA, Thiokol formulators replaced it with poly(butadiene-acrylonitrile-acrylic acid; PBAN), which was synthesized in a random, free radical polymerization almost identical to that for PBAA. The inclusion of the acrylonitrile improved the spacing between the acrylic acid groups (i.e., the crosslink sites), allowing for better elastomeric properties. PBAN also had less propensity to surface harden in comparison with PBAA, a problem caused by oxidative crosslinking of the unsaturation carbons, common to polybutadienes and their copolymers.[49]

Both the PBAA and PBAN prepolymers were synthesized by emulsifying the monomers in water using a quaternary ammonium salt as an emulsifier and azobisisobutyronitrile as the free radical initiator. Although the stoichiometry of the monomers could be varied over a wide range, typically the amount of acrylic acid was kept low enough that a 3000 Da molecular weight chain would nominally have only two carboxylic acid groups, with about 6% by weight cyano groups (thought to limit oxidative crosslinking of the olefins in the polymer backbone).

By any measure, PBAN was a successful prepolymer; more PBAN has been made and consumed than any other rocket motor binder. It was estimated that about 2.6 million kg of PBAN-based propellant was produced through 1997.[50] The large consumption was in part due to the sizes of the motors PBAN was used in, which included the massive Titan III/IV-A UA120 and Space Shuttle strap-on boosters.[32]

### G. CARBOXYL-TERMINATED POLYBUTADIENE

A needed improvement to PBAA and PBAN was ensuring that the carboxylate groups were spatially separated along the prepolymer chain, to improve elasticity. In the late 1950s, chemists at Thiokol synthesized carboxyl-terminated polybutadiene (CTPB) using free radical polymeriza-tion.[51] A dicarboxylic acid peroxide (usually glutaric acid peroxide) or azo dicarboxylic acid initiator was used to polymerize and terminate butadiene in a pressurized solution, with molecular weights around 3500–5000 Da.[49] Because this was an uncontrolled free radical polymerization, the prepolymer had a polydisperse molecular weight and was highly branched (similar to HTPB, see discussion below). The polymer produced using peroxide initiation was made by Thiokol (later Morton International, then Rohm & Haas) under the name HC-434; the azo-initiated polymer was produced by BF Goodrich under the Hycar name.

Anionic polymerization of butadiene reduced the polydispersity and eliminated branching. The synthesis was a complicated process, with the organolithium salt of methyl naphthalene used to make the initiator, the dilithium salt of isoprene. Once the desired molecular weight of the dilithium salt of polybutadiene was attained, the ends of the polymer chain were capped with carbon dioxide, followed by anhydrous hydrochloric acid to yield CTPB and lithium chloride as a by-product.[52] Anionically polymerized CTPB was produced by Phillips under the Butarez CTL name.

Although both CTPB and PBAN were successful materials, they relied on carboxylate chemistry for curing, which caused problems. The hydrogen bonding from the pendant carboxylates caused the prepolymers to have fairly high viscosities, complicating mixing, and the curing with multifunctional aziridines or epoxides was slow, sometimes tying up production equipment for weeks at a time. The aziridines could also rearrange to form oxazolines, which react far slower with carboxylic acids and could even homopolymerize in the presence of AP[49] (Scheme 2). The homopolymerization and oxazoline problems could be so bad, in fact, that 20–30% of the aziridines added to a mix did not contribute to curing at all. These problems were tolerated because of the belief that the aziridine homopolymer might form a shell around the AP, increasing adhesion between the binder and oxidizer, and because the oxazolines would eventually revert back to aziridines and cure the carboxylic acid groups. However, what generally resulted were ill-defined mixtures of both epoxide and aziridine curatives, chosen empirically based on the acidities and polarities of the mixes and the intuition of the formulators.

Notwithstanding these problems, CTPB formulations were ubiquitous in solid propellant formulations of the U.S. military throughout the 1960s. However, the curing problems were so prevalent that within a decade, most newly qualified rocket motors were using polyurethane chemistry. Legacy systems continued with CTPB, owing mostly to the difficulty and cost of requalifying old motors for new ingredients. However, the changeover was accelerated by a fire in 1996 at Phillips's Butarez plant in Borger, Texas. Phillips declined to rebuild the plant, ending production of linear CTPB, and thereby leaving a number of U.S. missile programs without a major ingredient. Although attempts were made to replace Butarez with free radically-polymerized CTPB, in 2001 Goodrich was bought out by a group of investors, and Rohm & Haas announced an end to production of CTPB.

## H. HTPB

Hydroxyl-functionalized prepolymers had been combined with multifunctional isocyanates in the early 1950s, even before the introduction of PBAN and CTPB. General Tire & Rubber experimented with urethane crosslinked polyethers and polyesters for propulsion formulations as early as 1947. However, it was Aerojet that had the initial successes with these materials under the leadership of Karl Klager, an Austrian chemist who had worked for IG Farben during the Second World War and was brought to the United States by the U.S. Office of Naval Research under Operation Paperclip in 1949.[53,54] Klager's polyurethanes were used in both stages of the Polaris A1 and in the second stage of the Minuteman I. These materials were a combination of poly(1,2-propylene oxide) (PPG) and poly(1,4-tetramethylene oxide) (PTMEG), cured with toluene diisocyanate and crosslinked with triethanolamine. Other examples of such binders included

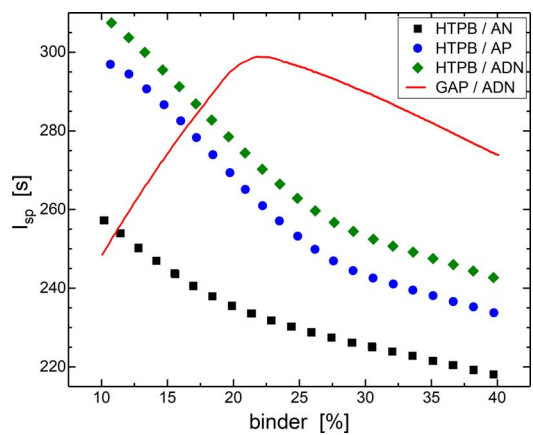

FIG. 4. — Specific impulse vs binder content for solid propellants employing HTPB with ammonium nitrate (squares), ammonium perchlorate (circles), ammonium dinitramide (diamonds), and an ammonium dinitramide formulation bound with glycidyl azide polymer (line). Adapted from ref 12.

poly(ethylene oxide) (PEG), poly(neopentylglycol azelate), and poly(butylene oxide); commonly referred to as B-2000. These types of binders had excellent stability, mechanical properties, and reliability for their era but nevertheless were passed over in favor of CTPB-based formulations.

So dominant was CTPB throughout the 1960s that although HTPB was first synthesized for use in binders in 1961, it did not see actual use in a rocket motor until 1968, when Aerojet used it in a dual-thrust radial-burning motor grain formulation for the Astrobee D, a NASA-sponsored meteorological sounding rocket.[50] Gradually, HTPB gained widespread use as a replacement binder in older systems employing CTPB, for example, in the Maverick, Stinger, and Sidewinder missiles. After 50 years of use, HTPB remains the standard binder for nearly all U.S.-made SRM. HTPB-based polyurethane binders are relatively inexpensive, have low viscosity prepolymers, and exhibit good mechanical and aging properties. The binary system enables a high solids content, providing one of the highest specific impulses among solid propellants. Figure 4 shows the $I_{sp}$ of HTPB-based solid propellants with various metals.[12] Notwithstanding its excellent properties, HTPB popularity no doubt benefitted from its being the most widely available prepolymer at the time SRM development activity largely ended in the United States.

The synthesis of the HTPB prepolymer is by free radical polymerization of butadiene in ethanol or isopropanol using hydrogen peroxide.[55] As such, it typically has a broad molecular weight distribution and substantial branching. Although there are nominally 2–3 hydroxyls per prepolymer chain, this is an average, and the chemical structure is polydisperse. Various studies of the hydroxyl group content reveal the presence of high-molecular-weight chains bearing several (sometimes more than a dozen) hydroxyl groups per chain, with the details varying from batch to batch.[56–58] This makes it more difficult to control the crosslinking of HTPB, in comparison with older systems based on anionically polymerized CTPB, in which a telechelic prepolymer bearing only two reactive groups was doped with a multifunctional crosslinking agent.

Curing of HTPB is typically carried out with an aliphatic diisocyanate, such as isophorone diisocyanate or hexamethylene diisocyanate. Generally, aromatic isocyanates are too reactive, shortening the pot life (toluene diisocyanate being an exception). When greater than two isocyanates per crosslinking agent are required, biuret triisocyanate is the typical choice. Although a catalyst is not necessary for urethane formation, often a Lewis acid catalyst (e.g., iron acetylacetonate, dibutyltin dilaurate, or bismuth trichloride) is used. Presumably, the metal catalyst coordinates to the isocyanate, lowering the activation energy for reaction with the hydroxyl.[59,60] The pot life of HTPB can be extended by fine-tuning of the polymer microstructure.[61]

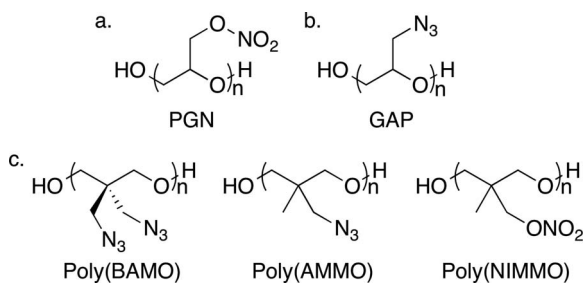

SCHEME 3. — Common energetic polymers.

Once Aerojet recognized HTPB's potential, production was established by Sinclair/ARCO under the trade name R-45M. This HTPB production facility, now owned by Total Petrochemicals, continues to operate.

## I. HYDROXYL-TERMINATED POLYETHERS

As noted above, polyurethane binders based on polyethers and polyesters had been successfully formulated into fielded missile systems in the early 1950s, superseded first by CTPB in the early 1970s and followed by HTPB thereafter. This was due in part to the fact that polybutadienes, when formulated in a standard motor, yield a slightly higher specific impulse than the polyethers (about a 3 s increase). Also, early polyethers tended to embrittle when exposed to humidity, which was practically unavoidable in a realistic setting. This caused problems with propellant aging and debonding of the motor to its liner. However, use of the first-generation polyethers continued in specialty applications, such as low-burning sustainer motors in dual-thrust rockets.[54]

In the late 1980s, the U.S. military began to emphasize reduced sensitivity for their ordnance; that is, a reduction of the violent effects of accidental propellant cook-off.[62] Greater insensitivity of the motor to impact could be achieved by reducing solids loading, but this reduces performance. The requirement for insensitive munitions thus precipitated a reevaluation of propellant binders to achieve better energetic performance to compensate for the lost performance. An obvious way to make a binder more energetic is to incorporate energetic groups on the prepolymer (see below). Alternatively, an energetic plasticizer could be used in the formulation. Energetic plasticizers such as N-butyl-N-nitratoethyl nitramine and trimethylolethane trinitrate are too polar to be soluble in polymers such as polybutadiene. Polyethers are an obvious solution to the solubility problem, particularly with the moisture embrittlement issue largely mitigated by judicious use of bonding agents.

Older polyethers, such PPG and PTMEG, were discussed above. A more recent example of a hydroxyl-terminated polyether binder is Terathane-PEG, a block copolymer of poly(1,4-butanediol; or Terathane) and poly(ethylene glycol) that was synthesized by DuPont and formulated by ATK for use with highly polar nitroplasticizers[63] in the early 1990s.

## VI. ENERGETIC POLYMERS

Under standard conditions, conventional solid propellants produce a specific impulse on the order of 265 s.[8] There are continuing efforts to develop binders that yield higher combustion energy, typically employing azides or nitrate esters. NC was the first energetic binder used in SRM, and there have been continuing efforts to develop more energetic binders. Invariably, these are hydroxyl-terminated prepolymers with urethane crosslinking.

### A. POLY(GLYCIDYL NITRATE)

In 1953, formulators at the U.S. Naval Ordnance Test Station in China Lake, California, synthesized 800 to 3400 Da poly(glycidyl nitrate; PGN) using stannic chloride as a catalyst,[64,65] although the generation of acetyl nitrate as a side product limited its scale-up potential at the time[43] (Scheme 3a). Eventually, glycidyl nitrate would be synthesized safely in a flow reactor with dinitrogen pentoxide with high yield and purity by the Defence Research Agency in the United Kingdom in the early 1990s.[66,67] Hydroxyl-functionalized prepolymer was prepared using a boron trifluoride catalyst initiator.[68]

### B. POLYOXETANES

Energetic polymers derived from oxetanes are polymerized using boron trifluoride chemistry, similar to PGN. These polyethers are based on 3,3-bis(azidomethyl) oxetane (BAMO), 3-azidomethyl 3-methyl oxetane (AMMO), and 3,3-(nitratomethyl) methyloxetane (NIMMO).[69] Naturally, the corresponding polymers (made by Aerojet but now discontinued) are known as poly(BAMO), poly(AMMO), and poly(NIMMO) (Scheme 3c) and have found limited use in small gas-generating rocket motors. Poly(BAMO) is the most energetic of the three but is used with either poly(AMMO) or poly(NIMMO) as a copolymer because of its higher crystallinity.

### C. GLYCIDYL AZIDE POLYMER

A chemically simpler polyether featuring the energetic azide group is glycidyl azide polymer, synthesized in the early 1970s[70] and evaluated as a possible binder starting in 1976[71] (Scheme 3b). In practice, glycidyl azide itself was found to be resistant to polymerization, so glycidyl azide polymer (GAP) was generated directly from the substitution of the chloride in polyepichlorohydrin by sodium azide in dimethylsulfoxide. GAP was evaluated by the U.S. Air Force as a propellant binder as early as 1981.

## VII. PROCESSING

Processing of solid propellants requires shaping and curing a highly loaded material with minimal heat buildup. A rule of thumb is that the uncured propellant, including solid oxidizer and metal particulate, have a viscosity no greater than *ca.* 50 Pa s in order to be successfully cast in a traditional mixer. This is much less than the viscosity of most binders without added plasticizers, which therefore are incorporated at levels of 10% or more by weight (with respect to the prepolymer). It is critical that the working time of the prepolymer is long enough to permit adequate mixing, casting, and curing under vacuum. In the case of a large SRM, such as a strap-on booster, several mixes, of up to 1600 L each, are transported inside the mixers to the casting pit and then poured directly into the motor case. Working time is regulated by temperature control and judicious use of catalysts for the crosslinking reaction.

Resonant acoustic mixing has been investigated as a replacement for traditional stirring.[72] This method uses a low-frequency, high-intensity acoustic field to effect mixing, with micro-mixing regimes generated throughout the precured "dough."[73] In theory, this leads to a more homogeneous mixture and allows the motor ingredients to be mixed directly in their case, obviating the need for separate mixing vessels.

A processing option that has not yet been implemented is additive manufacturing (three-dimensional printing) of the propellant. The advantages would include potentially more intimate mixing and the ability to form more complex grain structures, particularly within the center perforation, and thereby better tune the burning profile.[74,75]

## VIII. MECHANICAL PROPERTIES OF BINDER

The SRM grain is bonded to the case to immobilize the propellant and prevent premature burning of the outer surface of the grain. Debonding can be a significant problem, and additives are included in propellant formulations to improve bonding characteristics. Thermal stresses during storage are typically small (<0.1 MPa), although thermal cycling can induce fatigue cracks. More serious is shrinkage during the cure, which gives rise to stresses at the case boundary. Stress concentration in the vicinity of oxidizer particles, particularly large ones, can cause debonding from the matrix. Stress due to cure shrinkage can be compensated for by elevating the temperature somewhat above the cure temperature. Typical accelerated aging at 60–65 °C, corresponding to as much as 15 years at ambient temperature, was found to cause only modest changes in mechanical properties of HTPB-based propellants.[76,77] However, prolonged storage at low temperature can cause stresses to develop that result in dewetting of the binder from the oxidizer and consequent voids within the motor. Even relatively small static preloads prior to ignition can initiate void formation.[78] Expansion of a small void can lead to catastrophic failure (bursting) of the SRM, if the developing hole reaches a critical size.[79–82] The largest stresses are encountered during ignition and flight, with typical magnitudes of *ca*. 0.5 MPa. Although this is well below the strength of the materials, the stresses are sufficient to cause substantial dewetting of the binder[83] and loss of grain structural integrity.[84]

Accumulation of interfacial cracks and holes during cure, storage, and operation affects the mechanical properties and structural stability of the propellant and ultimately can lead to catastrophic failure. Thus, the design of solid propellants is guided by analysis and prediction of the stresses exerted on the binder–grain interface, particle–binder debonding, and the development of voids. An accurate assessment of damage tolerance is critical. Experiments and modeling are used to deduce the softening of the propellant due to damage, with the assumption usually made that any softening arises due to void formation; that is, mechanical hysteresis due to viscoelasticity or anelastic effects is usually neglected.

## IX. MODELING

Modeling the mechanical behavior of a solid propellant is essential for rational design of SRM and to ensure both their performance and safe operation, the latter including estimation of service lifetimes. However, this modeling is challenging, as it entails virtually all factors that complicate such analyses: nonlinear strains, mixed strain modes, high strain rates, failure and fracture of the material, and temperatures and stresses that change over time. The material itself is complex, involving multiple components that have mechanical properties that are very different yet coupled. The initial step is to derive a constitutive equation, which quantifies the relationship of stress to strain, including their evolution over time. For polymers, elasticity models are used, which can be phenomenological or molecular. The former typically describe the mechanical response in terms of derivatives of the strain energy, to arrive at tractable equations that, however, have no connection to the chain molecules.[85–87] The strain energy is usually expressed as a series expansion in terms of strain invariants (strain descriptors that are the same for any orthonormal coordinate system used to represent the strain components), with the linear terms corresponding to ideal elasticity and the well-known Mooney–Rivlin equation.[85,88] Nonlinear terms can be added to improve the fitting of experimental data, although this can lead to large errors on extrapolation. Vahapoglu and Karadeniz[89] reviewed the various phenomenological equations through 2003. Molecular models of rubber elasticity,[88,90,91] based on chain entropy, have a degree of accuracy similar to phenomenological approaches but with the advantage of providing insight into structure–property relations.

The effect of filler reinforcement is an added complication of modeling efforts. There are a surfeit of models for the viscosity of fluids that are highly filled with particles of varying size.[92–96] The basic concept is straightforward: strain amplification due to the inextensibility of hard particles enhances the stiffness (albeit limited by chain scission or detachment from the filler particles).[97] That is, the strain energy arises from rubber elasticity, amplified by the presence of filler. The role of the polymer–filler interface, including bound and occluded rubber, is significant. Moreover, particle deagglomeration and slippage of interfacial chains impart a strain dependence, which can dominate the filler effect at large strains.

A polydisperse particulate (Figure 3) is a further complication, although even generic carbon black exists in rubber over a size range that can span from 10 to $10^4$ nm.[98] The number of expressions describing reinforcement reflects this dispersity. For discretely sized particles, these have the general form of a product series

$$\eta(\phi) = \eta_{\phi=0} \prod_{i=1}^{n} H(\phi_i) \quad (2)$$

in which $\phi$ is the filler volume fraction and each factor corresponds to a discrete mode (bin) in the size distribution.

The stiffening function, $H(\phi)$, can include the effects of particle shape and interactions. The starting point is the Einstein equation, valid for low particle concentrations:

$$\eta = \eta_{\phi=0}(1 + 2.5\phi) \quad (3)$$

Typically, for carbon black reinforcement, a modification is the Guth–Gold equation in which $\eta_{\phi=0}$ is the viscosity in the absence of filler:[99]

$$\eta = \eta_{\phi=0}(1 + 2.5\phi + 14.1\phi^2) \quad (4)$$

Commonly, $\phi$ is taken to be an effective volume fraction, larger than the actual value due to occluded rubber.[100,101] For higher concentrations of monodisperse particles, many equations have been proposed, most of which are empirical. Examples that have been found to represent experimental data well include[102]

$$\eta = \eta_{\phi=0}\left(1 + 2.5\phi + \kappa\phi\left[\frac{\phi}{\phi_m - \phi}\right]^2\right) \quad (5)$$

in which $\kappa$ is an adjustable parameter, and[103]

$$\eta = \eta_{\phi=0}\left(1 + 0.75\frac{\phi/\phi_m}{1 - \phi/\phi_m}\right) \quad (6)$$

which reduces to the Einstein equation for a maximum volume fraction of solids $\phi_m = 0.605$. Given the correspondence between the viscosity and modulus, $\eta/\eta_{\phi=0} = E/E_{\phi=0}$ (for the tensile modulus), Eqs. 2–6 are also used for the effect of hard particles on the modulus.

Common assumptions in modeling the mechanics of solid propellants are (i) reversible deformation without permanent set; (ii) no volume changes induced by strain; (iii) an absence of strain localization, such as necking; and (iv) the effects of strain and time (or strain rate) are uncoupled. Propellants are susceptible to flow, and their viscoelastic nature must be included in the constitutive modeling. Very generally in modeling rubber, the stress is expressed as a simple product of functions of time and strain. Using some form of the Boltzmann superposition principle, for uniaxial tension the stress is

$$\sigma(t) = \int_0^t \frac{dE(t-u)}{du}\Big(\varepsilon(t) - \varepsilon(u)\Big)du + E(t)\varepsilon(t) \qquad (7)$$

in which $\varepsilon$ is the tensile strain. The first term in the integral represents the stress remaining at time $t$ of stress that arose at time $u$. The second term is the isochronal stress. Rubber is arguably the most "linear" of materials, at least using an appropriate definition of $E(t)$.[88,104] With experimental characterization of the strain and rate dependences, this approach works reasonably well for nonreversing deformations. However, when the sign of the strain is changed, Eq. 7 and similar treatments invariably underestimate the amount of energy dissipated. Both phenomenological and molecular constitutive equations share this failure to describe reversing strain histories of rubbery networks.[105,106] This means that when constitutive parameters are obtained by fitting experimental tension or shear data, the stresses predicted during recovery are too large; that is, the measured hysteresis exceeds the calculated energy dissipation.[107–109] There are recent efforts to incorporate anelastic features into rubber elasticity models,[110,111] although very generally, the mechanical behavior of rubber networks during recovery remains poorly understood.[112–114]

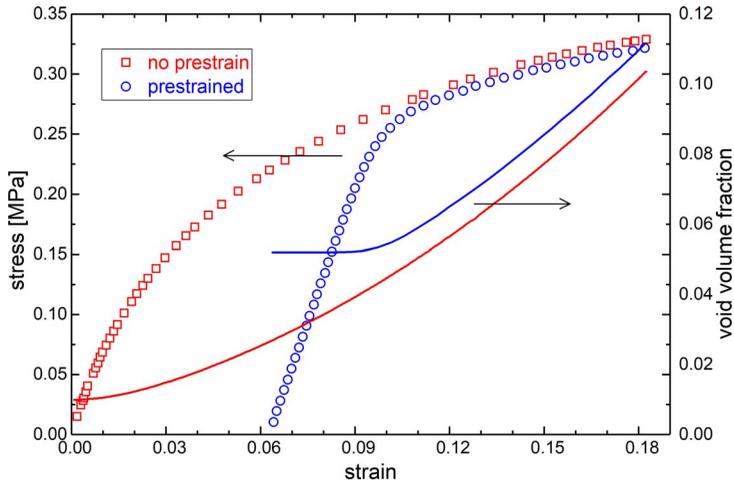

FIG. 5. — Development of voids, as predicted from a model ascribing softening to debonding of the binder from the particles. From ref 78.

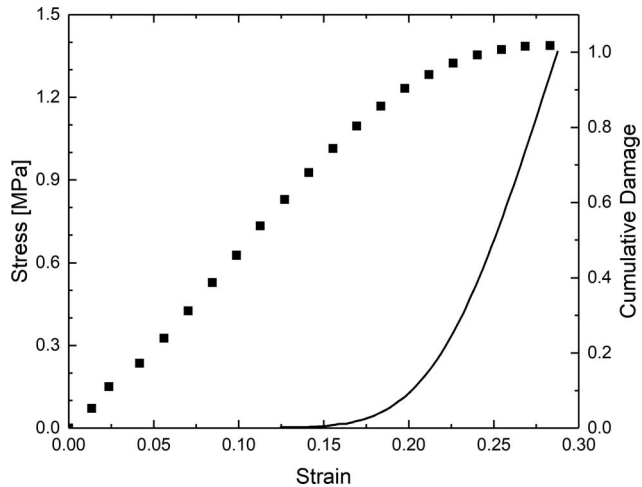

FIG. 6. — Stress (symbols) and cumulative damage (line) for an HTPB solid propellant. From ref 15.

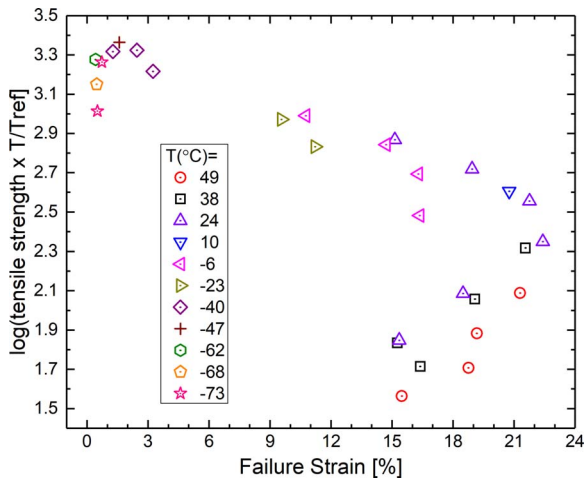

FIG. 7. — Failure envelope for a typical solid rocket propellant.[131]

## X. BINDER FAILURE

Constitutive equations for solid propellants usually include empirical damage terms to quantify the softening of the propellant, ascribing it to damage accumulation. The damage material parameters in these constitutive equations are determined from the deviation of experimental stress–strain measurements; that is, an empirical softening function is adjusted to agree with the experimental data.[78,115–121] These models yield accurate stresses during initial deformation of the propellant; however, the predicted recovery stresses are too large.[122–124] The problem is when modeling solid propellants, the failure of expressions such as Eq. 7 for reversing strains is neglected, and thus the "excess" energy loss attributed to structural damage such as dewetting and void formation is overestimated.

Damage models often treat the underlying mechanism as a transition of filler particles to voids, based on the assumption that debonded particles exert no reinforcement.[78,79,116,119,120] The strength of the interfacial adhesion between particle and matrix is assumed to be a material constant, independent of the deformation conditions.[125–127] Figure 5 shows results for the void volume calculated for a typical solid propellant, deduced from the deviation of the stress from values calculated from the constitutive equation for the propellant. Cumulative damage obtained by an analysis of this type is shown in Figure 6.[115]

The intractable problem in modeling elastomers in general and solid propellants in particular is separating irreversible structural changes such as void formation from the viscoelastic hysteresis intrinsic to an amorphous polymer above its glass transition temperature.[128] A common misperception is that this softening, referred to as the Mullins effect, is especially substantial in highly filled elastomers. This would include propellant binders, and indeed, they exhibit marked mechanical hysteresis. However, all viscoelastic materials exhibit Mullins softening,[106] which in rubber can exceed 20% of the strain energy even at low strain rates ($<10^{-3}$ $s^{-1}$).[105] In fact, the magnitude of Mullins softening is very similar for filled and unfilled compounds when compared at the same peak stresses.[128,129] Because of the inherent limitations of the constitutive equations for rubber, and because values for the damage parameters must be obtained by fitting experimental data, the models for propellants lack predictive capability.

TABLE III
TYPICAL MATERIAL PROPERTIES OF SOLID ROCKET MOTOR COMPONENTS[136,137]

|  | Propellant | Steel case | Insulation |
|---|---|---|---|
| Modulus, MPa | ~1 | $210 \times 10^3$ | 11 |
| Poisson's ratio | 0.5 | 0.3 | 0.3 |
| Thermal expansion coefficient, $K^{-1}$ | $1.1 \times 10^{-4}$ | $0.11 \times 10^{-4}$ | $2.3 \times 10^{-4}$ |
| Thermal conductivity, W/($m^2$ K) | 0.61 | 42.5 | 0.36 |
| Heat capacity, J/(g K) | 0.83 | 0.46 | 0.36 |

The range of strains, pressures, and temperatures experienced by solid propellants are very broad, making it difficult to determine the failure limits of a binder from laboratory measurements. One approach used for solid propellants is to develop failure envelopes,[130,131] based on the early work of Smith.[132,133] In this method, the stress at break is plotted versus the failure strain, with values obtained at different temperatures and strain rates presumed to fall on a single curve. This curve, the failure envelope, defines the mechanical limits of the material for arbitrary conditions. The approach assumes (incorrectly[134,135]) that time–temperature superpositioning is valid over the range from the low-frequency polymer chain dynamics to fast local segmental motions. Shown in Figure 7[131] is the failure envelope for a typical solid propellant, which ideally would define the safe operating range of the material.

## XI. AGING

An important consideration in propellant performance is aging, which can entail (i) chemical changes in the binder, affecting the modulus and combustion behavior; (ii) crystallization of some components; (iii) dewetting and porosity development; (iv) phase separation of components within the grain; (v) moisture ingress, affecting modulus and burn characteristics; and (vi) separation of the grain from the liner. Increases in stiffness due to crosslinking during storage can induce cracking.[136] This cracking is exacerbated by the very different material properties of the components of a rocket motor (Table III).[137,138] These differences include both thermal properties and consequent thermal loads as well as differences in component stiffness, which amplify vibrational perturbations. Oxygen diffusion through the rocket motor grain is quite high, and oxidative damage once the antioxidant is exhausted can be extensive, especially at elevated storage temperatures.[139,140]

## XII. CONCLUDING REMARKS

Their low cost, long shelf life, and immediate readiness ensure that solid propellants will remain in wide use for both military and civilian applications.[141] Although the performance of solid propellants is significantly lower in comparison with liquid fuels, SRM thrust profiles are predictable and achieved with relatively small volumes. Innovative designs even provide the ability to stop and restart an SRM after ignition. Present development work is focused on increasing energy yields, in response to the need to reduce the solids content in order to meet insensitive munitions requirements and to improve the reproducibility and efficiency of processing. The latter includes exploring additive manufacturing, which offers the potential for more control of grain geometries. There is no lack of modeling efforts that address the failure of composite binders; however, these efforts suffer from the general limitations of rubber modeling, most evident when compounds are subjected to reversing strain histories.